# Evidence of a universal and isotropic $2\Delta/k_B T_C$ ratio in 122-type iron pnictide superconductors over a wide doping range

*Xiaohang Zhang,[1,*] Yoon Seok Oh,[2] Yong Liu,[2] Liqin Yan,[2] Shanta R. Saha,[1] Nicholas P. Butch,[1] Kevin Kirshenbaum,[1] Kee Hoon Kim,[2] Johnpierre Paglione,[1] Richard L. Greene,[1] and Ichiro Takeuchi[3]*

[1]CNAM and Department of Physics, University of Maryland, College Park, Maryland 20742, USA

[2]CeNSCMR, Department of Physics and Astronomy, Seoul National University, Seoul 151-747, South Korea

[3]Department of Materials Science and Engineering, University of Maryland, College Park, Maryland 20742, USA

(Dated on July 29, 2009)

**ABSTRACT** We have systematically investigated the doping and the directional dependence of the gap structure in the 122-type iron pnictide superconductors by point contact Andreev reflection spectroscopy. The studies were performed on single crystals of $Ba_{1-x}K_xFe_2As_2$ ($x$ = 0.29, 0.49, and 0.77) and $SrFe_{1.74}Co_{0.26}As_2$ with a sharp tip of Pb or Au pressed along the $c$-axis or the $ab$-plane direction. The conductance spectra obtained on highly transparent contacts clearly show evidence of a robust superconducting gap. The normalized curves can be well described by the Blonder-Tinkham-Klapwijk model with a lifetime broadening. The determined gap value scales very well with the transition temperature, giving the $2\Delta/k_B T_C$ value of ~ 3.1. The results suggest the presence of a universal coupling behavior in this class of iron pnictides over a broad doping range and independent of the sign of the doping. Moreover, conductance spectra obtained on *c*-axis junctions and *ab*-plane junctions indicate that the observed gap is isotropic in these superconductors.

[*] Email address: xhzhang@umd.edu

The discovery of superconductivity in different classes of iron pnictides[1] has attracted great interest. To date, a widely discussed pairing theory for the superconductivity is the $s\pm$-wave symmetry[2] with nodeless gaps for both electron pockets and hole pockets and a sign reversal between them. Experimentally, a number of observations[3-17] have been suggested to be consistent with the $s\pm$-wave picture; however, there remain significant discrepancies in the reported results. First of all, at present there is no consensus on the value of the $2\Delta/k_B T_C$ ratio,[3-21] which represents the fundamental coupling strength of superconductivity. Moreover, it has been suggested that many experimental data can be explained with either the $s\pm$ wave or a single gap picture,[10-12,14,17-21] implying a lack of conclusive evidence for the multi-gap prediction of the $s\pm$ wave in these measurements. Perhaps, the most intriguing observation comes from recent thermal transport measurements, in which a substantial residual $\kappa_0/T$ term at zero field was found in $BaFe_2As_{2-x}P_x$ (Ref. 22) and $KFe_2As_2$ (Ref. 23) single crystals, indicating the existence of line nodes in the energy gap of these superconductors. This particular result has lead to the speculation that perhaps the structure of the order parameter changes upon doping from nodeless to nodal. Clearly, a systematic doping dependent study of the order parameter is required to understand the pairing nature in the 122 type superconductors.

As direct probes of the density of states, tunneling spectroscopy and point contact Andreev reflection (PCAR) spectroscopy have historically played central roles in investigating the superconducting gap. In addition to providing the measurement of the $2\Delta/k_B T_C$ ratio, the ability of these techniques to study the anisotropy and the temperature dependence of the gaps makes them a key tool in providing direct evidence for various mechanisms of superconductivity.[24]

A quick survey of the only few tunneling and PCAR measurements conducted on the 122-type pnictide superconductors to date reveals a wild variation in the reported gap values. Scanning tunneling microscopy (STM) measurements on potassium doped[19] and cobalt doped[20,21] pnictides indicate a spatially dependent single gap with an average $2\Delta/k_BT_C$ ratio of 6-7 estimated from the two coherence peaks. The subgap spectra obtained in these STM studies are typically V-shaped, inconsistent with the BCS *s*-wave superconductivity. An early *ab*-plane direction PCAR experiment by Szabó *et al.*[16] suggested a two-gap structure in $Ba_{0.55}K_{0.45}Fe_2As_2$ with $2\Delta/k_BT_C$ ratios of 2.5-4 and 9-10; however, no superconducting gap features were observed in the *c*-axis direction. A recent PCAR study on $Ba_{0.6}K_{0.4}Fe_2As_2$ (Ref. 18) indicates a single superconducting gap with a $2\Delta/k_BT_C$ ratio of 2.0-2.6. Moreover, measurements on cobalt doped 122-type crystals have also led to quite different observations: in the absence of a clear feature for the superconducting gap, a large zero-bias conductance peak was observed by Lu *et al.*[18] while a single gap with a $2\Delta/k_BT_C$ ratio of ~ 5.0 was found by Samuely *et al.*[17] The inconsistencies in these results most likely arise from a perennial issue in PCAR, namely, non-ideal surface and interface conditions.[25,26] Moreover, the ballistic nature of the PCAR transport requires high-transparency contacts to restrain the spectrum broadening effect.

In this work, we use high-transparency contacts to perform a systematic and consistent PCAR spectroscopy study on both hole-doped (K doped) and electron-doped (Co doped) 122-type iron pnictide superconductors over a wide doping range. The conductance spectra show only a single robust superconducting gap, and the gap value can be precisely determined by using the Blonder-Tinkham-Klapwijk (BTK) model[27] with a lifetime broadening term.[28] From the obtained gap values on various crystals in this class of pnictides, a clear picture emerges: the superconducting gap is isotropic with a constant coupling strength $2\Delta/k_BT_C$ of ~ 3.1 over the entire doping range studied here.

Four batches of single crystals grown in FeAs-flux (for Co-doped)[29] or Sn-flux (for K-doped)[30] were used in this study. Bulk superconductivity of the crystals was confirmed by magnetic susceptibility measurements. Wavelength dispersive x-ray spectroscopy and energy dispersive x-ray spectroscopy results showed that the exact compositions of the crystals were $Ba_{0.71}K_{0.29}Fe_2As_2$, $Ba_{0.51}K_{0.49}Fe_2As_2$, $Ba_{0.23}K_{0.77}Fe_2As_2$, and $SrFe_{1.74}Co_{0.26}As_2$. Resistivity measurements indicated the superconducting transition temperatures of the crystals are 28 K, 25.5 K, 21 K, and 15.5 K, respectively, with relatively narrow transition width of $\Delta T_C < 1.0$ K (estimated from 10-90% of normal state resistivity). More characteristic data of our crystals can be found in Refs. 30 and 31.

Point contact junctions were made on the crystals with sharpened tips of Pb and Au pressed along the *c*-axis or the *ab*-plane direction. The detailed junction formation procedure has been described elsewhere.[26] The advantage of using Pb as the counterelectrode in this study is that the observation of the weak-link type Josephson coupling [Fig. 1a] at low temperatures can be used as an indication of the existence of a highly transparent junction interface. Our previous analysis of the observed Josephson effect in Pb/$Ba_{1-x}K_xFe_2As_2$ junctions[26] has indicated that the current across the interface flows through many parallel Sharvin-type channels within a typical contact area of ~ $10 \times 10$ (μm)$^2$. Specifically, our picture is that there is formation of a complex distribution of a dead layer at the crystal surface, and combination of this surface with the detailed nanostructure of the tip leads to multiple transparent contacts at the interface with a contact scale of each channel smaller than the mean free path of the quasi-particles which is on the order of 10 nm.[20] This picture is also consistent with the observation of AR at the junction interface in the voltage range where quasi-particle transport is dominant [Fig. 1b]. Conductance spectra of weak-link type Josephson junctions have previously been used to study the gap value of superconductors through AR.[32] In our measurements, to avoid potential artificial features widely observed in PCAR measurements as described by Chen *et al.*,[33] conductance spectroscopy measurements were only pursued on junctions which displayed Josephson current (for Pb tip) or a junction resistance between 0.5 Ω and several ohms at finite bias voltages (for Au tip).

Conductance spectra were measured via a phase-sensitive detection at a frequency of 173 Hz with a lock-in amplifier and a dc voltage source. For each junction, spectra were measured at temperatures from 4.2 K to the $T_C$ of the crystal with an increment of 1 K or 2 K. Figure 1b shows the normalized conductance spectrum at 8 K obtained from a *c*-axis Pb/Ba$_{0.23}$K$_{0.77}$Fe$_2$As$_2$ junction, which displayed a resistively shunted Josephson junction characteristic at low temperatures [Fig. 1a]. A conductance enhancement, with a ratio close to 2, appears at ~ ±3 meV confirming that the contact is highly transparent. To quantitatively describe the conductance spectrum and resolve the gap value, a modified BTK model is applied, in which three parameters are introduced: the superconducting gap Δ; a dimensionless parameter Z, which represents the interface transparency;[27] and an imaginary quasiparticle energy modification γ,[28] which reflects the spectral broadening. The best fit to the spectrum [Fig. 1b] results in a gap value of 2.7 meV. For a highly transparent junction (Z → 0), the normalized AR conductance enhancement is typically greater than 1.5 at low temperatures, and the lifetime broadening term is small, which would significantly reduce the fitting error for the energy gap. In our measurements, the fitting parameter Z is consistently less than 0.5 while the γ term is usually less than 1.0 meV, limiting the gap value uncertainty to about 0.2 meV at low temperatures.

Figure 2 shows differential resistance/conductance data and their analysis for representative junctions we studied. Figures 2a, 2c, and 2e are results from a *c*-axis Pb/Ba$_{0.51}$K$_{0.49}$Fe$_2$As$_2$ junction, while Figs. 2b, 2d, and 2f are from a *c*-axis Au/Ba$_{0.71}$K$_{0.29}$Fe$_2$As$_2$ junction. Raw differential resistance spectra of the junctions at 4.2 K are shown in Figs. 2a and 2b. In each case, a parabolic fit (dashed lines) to high bias data, consistent with a voltage sweep at a temperature slightly above the $T_C$ of the crystal, was used to normalize the spectrum.[34] The normalized conductance spectra of the junctions at selected temperatures are shown in Figs. 2c and 2d. At 4.2 K, the Josephson current is clearly evident in the Pb/Ba$_{0.51}$K$_{0.49}$Fe$_2$As$_2$ junction as a sharp peak at zero bias. In the low bias voltage range, the spectrum of the junction exhibits a symmetric conductance enhancement without any subgap feature due to multiple AR processes which have been previously reported.[32] The conductance enhancement shows negligible changes from 4.2 K to 8.0 K as the temperature is increased to slightly above the $T_C$ of Pb. Thus, the enhancement is truly due to the superconducting gap of the single crystal. As shown in Figs. 2b and 2d, when appropriately low resistance junctions are formed, similar features and a clear gap structure are also attained for junctions with an Au tip.

To evaluate the gap value for the crystals, the normalized conductance spectra were individually fitted by the modified BTK model mentioned above. As shown in Figs. 2e and 2f, the fitting parameter Z remains a relatively small value below the $T_C$ of the crystal for both junctions, indicating the high stability and transparency of the contacts. The broadening term γ is typically less than 1 meV with a slight variation in the temperature range below $T_C$. Although the reason for such a variation has not been unambiguously established,[34] the small lifetime broadening term does significantly increase the accuracy of the estimated gap size.

We have also applied the same analysis to the conductance spectra obtained on other junctions. An underlying feature of all spectra is that only a single gap is *unambiguously* identified. This observation is consistent with previous STM results[19-21] and a recent PCAR study.[18] However, within a double-gap picture, a possible explanation is that the values of the two gaps are too close to each other, and thus higher resolution or momentum separation methods are required to distinguish the two. This possibility has been proposed in a previous PCAR report on BaFe$_{1.86}$Co$_{0.14}$As$_2$ single crystals.[17] Alternatively, for AR of a double-band superconductor, if the corresponding feature of a gap is intrinsically small as discussed by Golubov *et al.*,[35] or if one type of carriers dominates the transport at the interface [as indicated by previous Hall effect measurements at temperatures close to $T_C$ (Refs. 31 and 36)], it would be possible that some features are smeared out and only a single robust gap is present in AR/tunneling spectra within the resolution of the measurements. In fact, we have occasionally observed extra features in our spectra at slightly high biases above the gap [as indicated by arrows in Fig. 2c], which may be an indication of the second gap[35] or related to AR bound states due to the s± symmetry.[37] However these features, compared to the main conductance enhancement associated with the predominant gap, were always very small in magnitude and were not always present. This precludes us from drawing a definitive conclusion as to their origin. Moreover, as previously reported,[33] such features could also be due to other artifacts. Therefore, in the present work, we have not addressed these small features.

Figure 3a shows the temperature dependence of the reduced gap as a function of the reduced transition temperature for four representative junctions made on the four respective types of single crystals. The uncertainty of the gap value increases with rising temperature due to the increase in the $\gamma/\Delta$ ratio. Within these uncertainties, the obtained data agree well with the BCS theory which is indicated by the dashed line. For junctions made on the same batch of crystals, the determined gap values are highly consistent with each other within an experimental error. Figure 3b shows the low-temperature gap value obtained on each junction as a function of the transition temperature of the single crystal. A linear fit to the data through origin can be obtained with the $2\Delta/k_B T_C$ ratio of approximately 3.1, suggesting that there is a constant coupling strength in the 122-type iron pnictide superconductors over a large doping range and independent of the doping type. As mentioned above, due to the high transparency of the junctions formed here, the uncertainty of the obtained $2\Delta/k_B T_C$ ratio is small. This coupling strength is slightly lower than the BCS weak-coupling value for an *s*-wave superconductor and comparable to the values obtained in previous PCAR measurements.[16,18] However this $2\Delta/k_B T_C$ ratio is much smaller than that estimated directly from the coherence peaks in STM measurements.[19-21]

In order to probe possible anisotropy (*c* axis vs *ab* plane) of the gap, junctions were also made by pressing tips onto side surfaces of both K-doped and Co-doped single crystals. As an example, Fig. 4 shows the normalized conductance spectra and the corresponding fits for a *c*-axis junction [Fig. 4a] and an *ab*-plane junction [Fig. 4b] fabricated on the same piece of $SrFe_{1.74}Co_{0.26}As_2$ crystal. We find that the gap values obtained for both types of junctions are approximately the same for all crystals within an experimental error, which suggests that the energy gap is isotropic.

As discussed above, a recent thermal conductivity measurement[23] has indicated the presence of nodes in the fully doped $Ba_{1-x}K_xFe_2As_2$, i.e., $KFe_2As_2$. Assuming that nearly optimally doped (low doping concentration) compositions are nodeless, this observation points to a possible evolution in the gap structure as a function of doping concentration in this system. Our study shows that such a change does not take place for the doping level at least up to $x = 0.77$. Instead, our result indicates a universal pairing nature of the superconductivity, which is in agreement with recent specific heat measurements,[38] where the universal behavior is suggested by a striking scaling effect between the specific heat jump and the transition temperature over a wide doping range for the 122 type iron pnictide superconductors.

In summary, PCAR spectroscopy studies have been conducted on a series of the 122 type iron pnictide single crystals with various doping concentrations. The spectra obtained on highly transparent junctions clearly display a robust conductance enhancement from which a superconducting gap is identified based on the modified BTK model. Although some extra features were occasionally observed in our data which may point to the existence of multi-gap superconductivity or AR bound states, infrequent occurrence and the small signal of these features preclude us from drawing a definitive conclusion as to their origin at this time. The most important observation in this study is that for the determined predominant superconducting gap, its value scales well with the transition temperature, resulting in the $2\Delta/k_B T_C$ ratio of ~3.1 for all crystals. Moreover, our results on *c*-axis and *ab*-plane junctions imply a likely isotropic gap formation in these superconductors. The isotropic gap, together with the constant coupling strength provides a strong evidence for a universal pairing mechanism in the 122-type iron pnictides over a broad doping range.

The authors acknowledge fruitful discussions with I. I. Mazin, D. Parker, S. H. Pan and F. Wellstood. X.Z. would also like to thank K. Jin and P. Bach for technical help. Work at UMD was supported by the NSF under Grant No. DMR-0653535 and partly by the AFOSR under Grant No. MURI-FA95500910603; N.P.B. is supported by CNAM; I.T. is funded by NSF MRSEC at UMD (Grant No. DMR-0520471); and the Work at SNU was supported by national creative research initiatives and basic science grant (Grant No. 2009-0083512) by NRF.

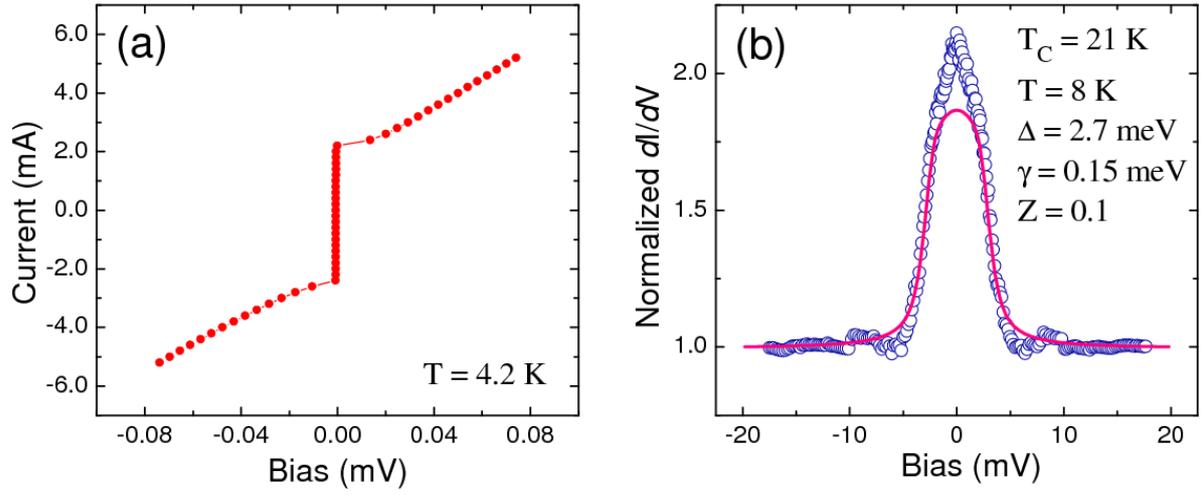

FIG. 1. (Color online) (a) A typical Josephson *I-V* characteristic at 4.2 K obtained on a *c*-axis Pb/Ba$_{0.23}$K$_{0.77}$Fe$_2$As$_2$ junction and (b) normalized conductance spectrum (circles) obtained on the same junction at 8 K with a modified BTK fit (line).

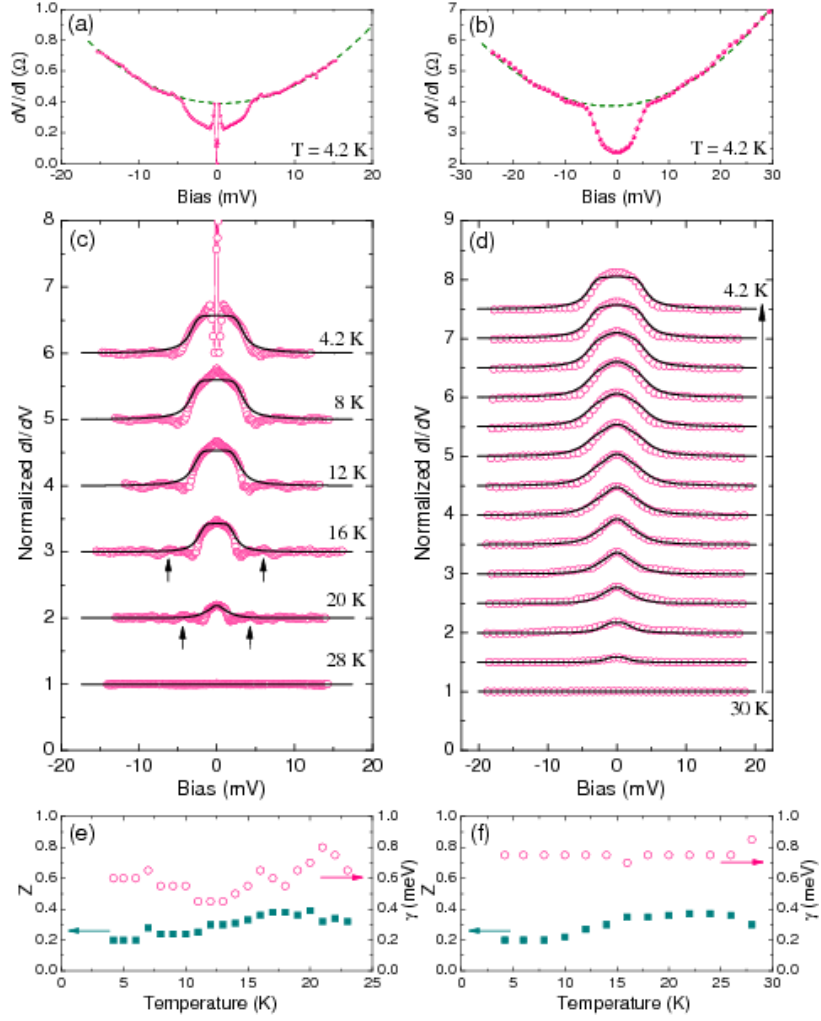

FIG. 2. (Color online) Differential resistance/conductance spectra and analyses obtained on [(a), (c), and (e)] a $c$-axis Pb/Ba$_{0.51}$K$_{0.49}$Fe$_2$As$_2$ junction and on [(b), (d), and (f)] a $c$-axis Au/Ba$_{0.71}$K$_{0.29}$Fe$_2$As$_2$ junction. (a) and (b) show examples of spectrum normalization as described in the text. (c) and (d) are temperature dependences of the normalized resistance spectra (circles) with BTK fits (lines). Data are vertically shifted for clarity except the bottom ones. (e) and (f) show the temperature dependence of Z and $\gamma$ used in the fittings. Arrows in (c) indicate small features as discussed in the text.

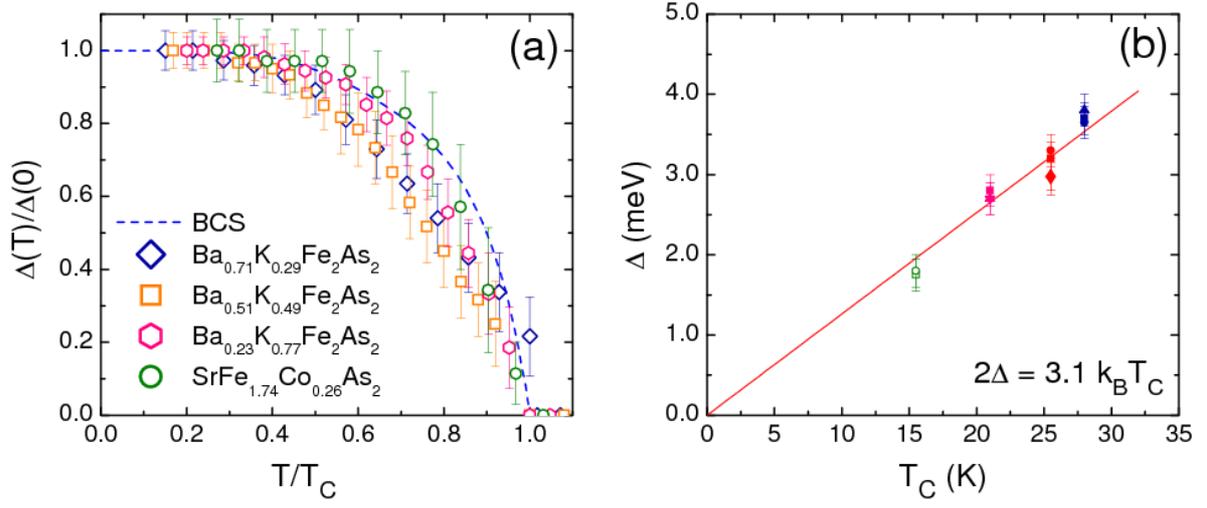

FIG. 3. (Color online) (a) Temperature dependence of the superconductor gap obtained on four representative junctions. Both gap value and temperature scales are reduced for each junction/crystal; the dashed line represents the BCS theory; (b) all low-temperature gap values obtained from our *c*-axis junctions are plotted as a function of the $T_C$. Solid symbols correspond to K-doped single crystals while empty symbols correspond to Co-doped samples. The solid line indicates a constant $2\Delta/k_B T_C$ ratio of ~3.1.

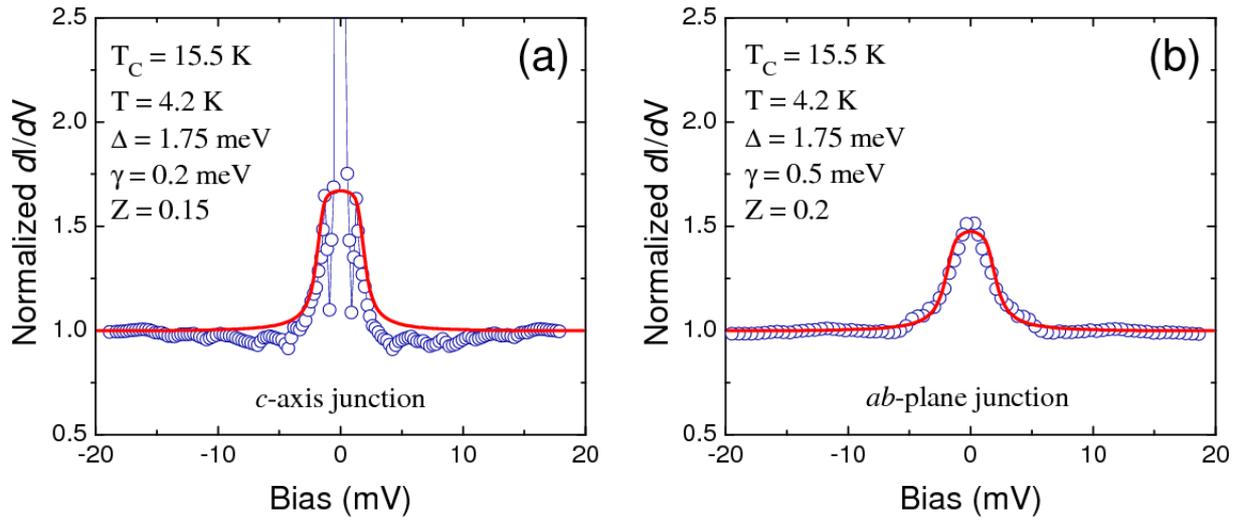

FIG. 4. (Color online) Normalized conductance spectra (circles) at 4.2 K obtained on (a) a $c$-axis Pb/SrFe$_{1.74}$Co$_{0.26}$As$_2$ junction and (b) an in-plane Au/SrFe$_{1.74}$Co$_{0.26}$As$_2$ junction. Lines are fits to the data.